# Experiments with calibrated digital sideband separating downconversion

Matthew A. Morgan - matt.morgan@nrao.edu
J. Richard Fisher - rfisher@nrao.edu

National Radio Astronomy Observatory
1180 Boxwood Estate Rd.
Charlottesville, VA. 22903

**Abstract**

This article reports on the first step in a focused program to re-optimize radio astronomy receiver architecture to better take advantage of the latest advancements in commercial digital technology. Specifically, an L-Band sideband-separating downconverter has been built using a combination of careful (but ultimately very simple) analog design and digital signal processing to achieve wideband downconversion of an RFI-rich frequency spectrum to baseband in a single mixing step, with a fixed-frequency Local Oscillator and stable sideband isolation exceeding 50 dB over a 12°C temperature range.

**1. Introduction**

Radio astronomy observations in the coming decade will require unprecedented levels of sensitivity while mapping large regions of space with much greater efficiency than is achieved with current telescopes. This requires new instrumentation with the greatest achievable sensitivity, dynamic range, and field of view (Webber & Pospieszalski 2002; Hall et al. 2008; Wilson et al. 2009). Receiver noise is quickly approaching fundamental limits at most radio wavelengths (Pospieszalski 2005), so significant gains in observing efficiency can be made only by increasing collecting area and fields of view. Jointly, these requirements suggest using large arrays of smaller antennas, or many moderate-size antennas equipped with multi-beam array feeds (Thompson et al. 2001; Emerson & Payne 1995).

The engineering community is thus challenged to develop receivers and wide bandwidth data transport systems which are lower cost, more compact, more reliable, lower weight, more reproducible, and more stable than the best current systems, *with no compromise to noise and bandwidth performance*.

This can be achieved with a greater degree of component integration, extensive use of digital signal processing and transport, and replacement of functions currently performed in expensive and bulky waveguide and coaxial cable components with digital arithmetic and thin optical fibers. There are no miracles to be pulled from the technological hat. All of this is to be performed with careful attention to detail and adoption of the latest products from the consumer and industrial electronics industry.

At the National Radio Astronomy Observatory, we have initiated a Research and Development effort to redesign and re-optimize receiver architecture to take full advantage of industry-driven technology. As well as transferring certain critical functions from the analog domain into the digital domain, this inevitably involves the seamless integration of the conversions from RF to baseband, from analog to digital, and from copper to fiber within a single receiver module. The result is a well-optimized modern receiver architecture that is compact, inexpensive, reliable, and manufacturable in relatively large quantities without compromising performance or versatility (Morgan & Fisher 2009).

**2. Digital Sideband-Separating Downconversion**

The first goal is to digitally sample the radio frequency (RF) signal as close to the antenna connection or focal point as possible. This reduces the total analog path length and amplifier gain required along with their associated temperature-dependant amplitude and phase fluctuations. Analog signal conditioning is still required to amplify the signal to the level required by the ADC and to limit the signal bandwidth to less than half of the digital sample rate to avoid sample aliasing.

The rms noise voltage at the input of a cryogenic receiver with a system temperature of 20 Kelvin and a bandwidth of 100 MHz is about 1 microvolt, whereas the ADC input signal level must be about 10 millivolts, so about 80 dB of net analog gain is required. Typical current receivers have 120 dB or more total gain to overcome multiple conversion, filter, and transmission line losses. A simplified receiver system will be more stable in at least two ways: less total gain and a much shorter signal path between the antenna terminals and the digitizer.

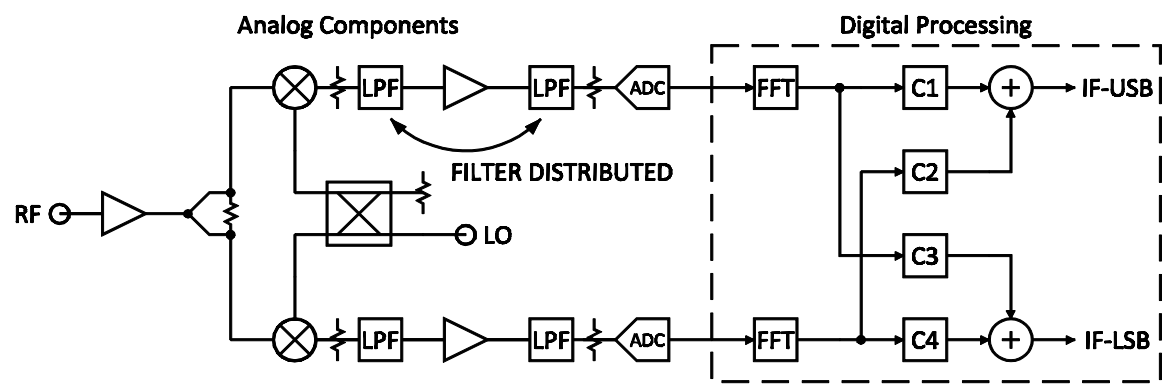

Figure 1. Block diagram of a Digital Sideband-Separating Mixer (DSSM).

Direct digitization of the RF signal without analog frequency conversion is a possible strategy (Vaughan et al. 1991), particularly at low to modest frequencies, but frequency tuning flexibility, strong signal rejection, and low power consumption often make analog conversion to a lower frequency band before digitization more attractive. It also allows the approximately 80 dB of gain to be divided between input and baseband frequency bands. Ensuring stability of 80 dB of gain at one frequency in a small space is a design problem one prefers to avoid.

Most heterodyne radio astronomy receivers incorporate at least two, often more, frequency conversion (mixing) stages to provide adequate frequency selectivity while tuning over broad bandwidths. Not only is this complex, but the use of multiple independent local oscillators (LO) opens the door for spurious mixing products. A cleaner solution is to use a sideband-separating mixer to go from RF to a near-zero intermediate frequency (baseband IF) in one step. A sideband-separating mixer requires a phase-quadrature power division of the LO signal and a broadband phase-quadrature combiner at IF (Maas 1993). The net phase- and magnitude-imbalance of analog power dividers and combiners typically limit sideband isolation to 20 dB or less, worse if the receiver must tune over a wide frequency range. This type of mixer is common at millimeter wavelengths (Chin et al. 2004; Kerr & Pan 1996), but poor sideband separation has ruled out its use in applications where radio frequency interference (RFI) is a problem or the astronomical spectrum is rich in spectral lines.

The solution is to move the IF signal combiner of a sideband-separating mixer into the digital domain. The I and Q output signals of the mixer pair, shown in Figure 1, are digitized separately and then arithmetically recombined using calibrated weighting coefficients. The weighting coefficients can be optimized not only to implement a mathematically perfect IF hybrid, but to compensate for phase and amplitude errors in the analog components. The ultimate performance depends only on the resolution of the signal processing arithmetic and the stability of the analog hardware. Compensated errors in analog components result in a small gain loss but no loss in signal-to-noise ratio.

This approach does not require a larger number of mixers or increase the digital data rate into the backend for a given total IF bandwidth. It merely replaces two different mixers in a multiple-downconversion scheme with two identical ones, and splits one analog to digital converter (ADC) into two ADC's with half the sample rate. Although the number of digital channels has been doubled, the total digital data rate for a given processed bandwidth is exactly the same (Fisher & Morgan 2008; Morgan & Fisher 2009).

### 3. Design Tradeoffs

There are a number of design tradeoffs to be considered in the context of a Digital Sideband Separating Mixer (DSSM) of the general form shown in Figure 1. The first concerns the distribution of gain in various stages of the receiver. Recall that approximately 80 dB net gain is required up to the ADC inputs. Additionally some amount of cryogenic gain is necessary to minimize the impact of the following stages on the total system noise. From a calibration stability standpoint, it is desirable to put as much gain as possible in front of the RF splitter, as gain fluctuations there will no have no effect on the sideband isolation. However, dynamic range considerations would tend to favor more gain on the IF side after down-conversion and filtering, affording more headroom to the mixers and potentially a reduction in the requisite LO power. Recall that putting too much gain at any one frequency increases the risk of instability in a compact integrated package. The optimum solution will depend on a number of variables including bandwidth, dynamic range requirements, and the performance of available components, but a reasonable place to start is to put about 30 dB of gain in the cryogenic portion, 30 dB in the RF path of the DSSM, and 30 dB in the IF path, allowing roughly 10 dB for downconversion and passive interconnect losses.

The filter passband shape is another important tradeoff to be considered carefully. Traditionally, anti-aliasing filters in radio astronomy have been high-order Chebyshev designs because the useful bandwidth (at least in terms of insertion loss) that they provide is closer to the Nyquist Limit than any other type of filter. However, the growth in digital technology and real-time processing capacity suggest that it is time to re-evaluate that selection.

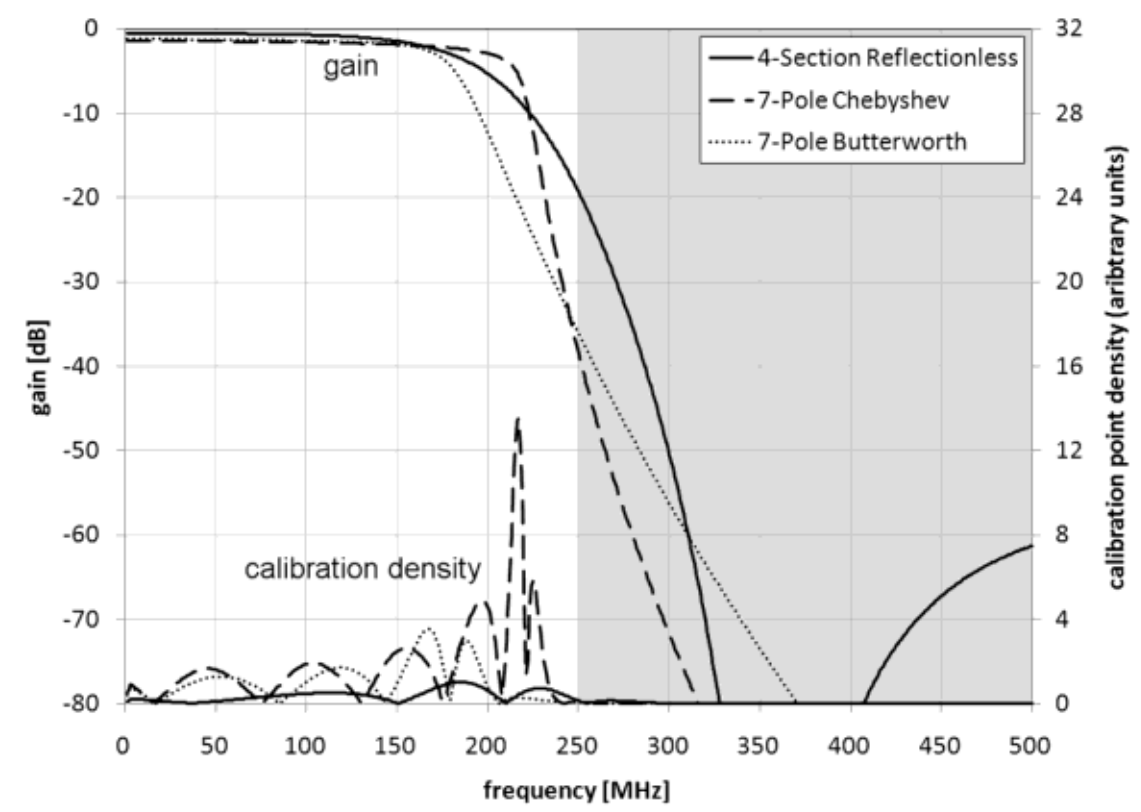

Figure 2. Passband characteristics and relative density of calibration points required for three choices of anti-aliasing filters -- a 4-section reflectionless filter, a 7-pole Chebyshev filter, and a 7-pole Butterworth filter. The inductor Q is 40 for all three filters. Each was tuned to provide 60 dB rejection of aliased signals throughout their 3 dB pass bands, assuming a sampling rate of 500 MS/s. The calibration density may also be interpreted as a figure of merit for the calibration stability, where larger calibration density implies poorer stability.

Figure 2 is a plot of three potential anti-aliasing filters that could be used in a DSSM. The first is a novel type of "reflectionless filter" that has theoretically infinite return loss at all frequencies in the passband, stopband, or transition band. This class of filter was developed specifically for use in the DSSM prototype to minimize standing wave interactions between the IF chain and the mixers at above-IF frequencies. In this frequency range, these filters may be implemented using standard off-the-shelf inductors, capacitors, and resistors. For this work, we have used surface-mount components, though in principle any fabrication technology may be used. What is important to the DSSM is that these filters do not reflect the stopband signals at either port, but instead absorb them. Their reflection coefficient is identically zero at all frequencies (in theory). This prevents the occurrence of standing waves between IF components.

Standing waves are problematic in at least two ways. First, if the electrical length of the transmission line supporting the standing wave changes (with temperature, for example), that change introduces a time-variable modulation of the signal amplitude. Second, the strong electromagnetic fields at the peaks and nodes of the standing wave enhance the coupling into the packaging cavity, which may be electrically large and thus exhibits cross-talk between adjacent channels which is very sensitive to temperature. The behavior of the reflectionless filters therefore help to improve the stability of the system as a whole. They are easy to implement, compact, and quite effective at eliminating problems associated with out-of-band reactive terminations (Morgan 2009). As we will show, they also have excellent complex gain characteristics themselves, which make it easy to achieve a precise, stable calibration.

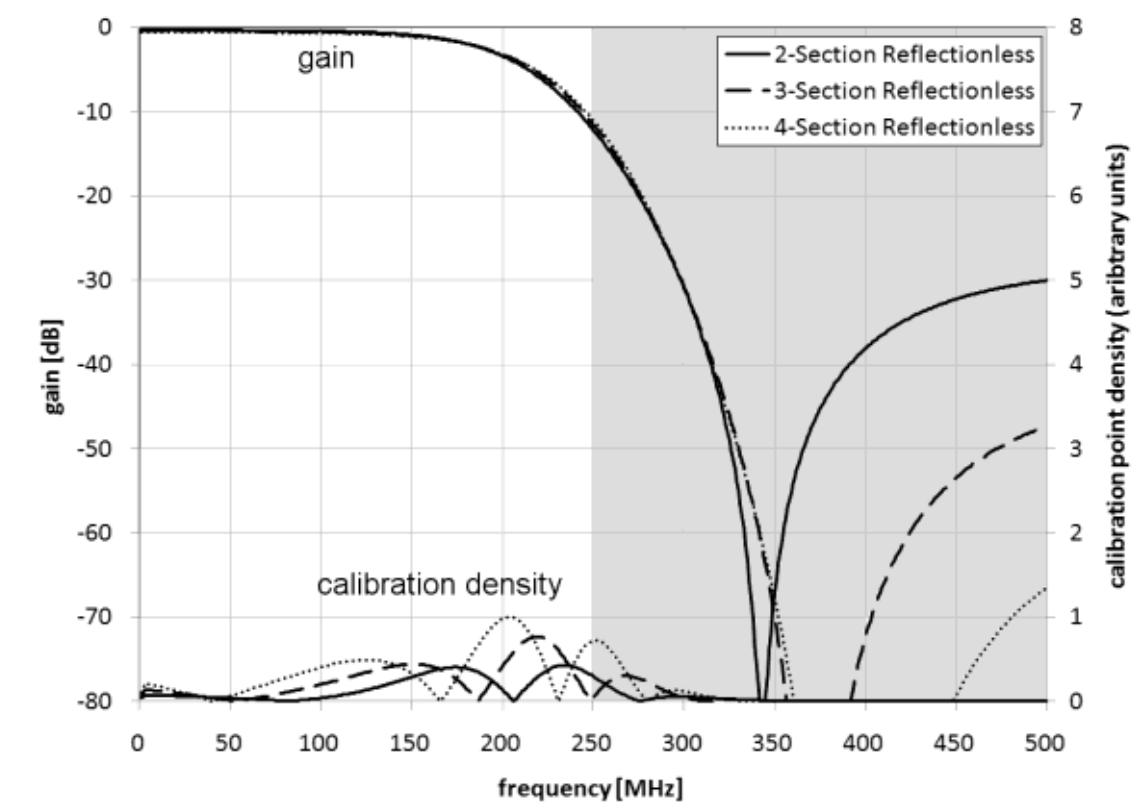

Figure 3. Passband shape and relative calibration density/instability as a function of filter order for the reflectionless filters. In this case, the alias rejection was held constant at 30 dB up to the filter's 3 dB passband, assuming a sampling rate of 500 MS/s.

The other two filters shown in Figure 2 are of more traditional topologies, namely a 7-pole Chebyshev filter, and a 7-pole Butterworth (or "Maximally-Flat") filter. To make a fair comparison, all three were designed for use in a system with sampling rate of 500 MS/s, wherein aliased signals are rejected by at least 60 dB up to the 3 dB cutoff of the passband. 60 dB isolation was selected for this experiment to be consistent with the level of sideband separation we hoped to achieve. In other words, we did not want aliasing to be the limiting factor in the suppression of nearby interferers. The inductor Q was assumed to be 40 in all three cases.

As expected, the Chebyshev filter has the greatest bandwidth, defined by its 3dB cutoff point. However in this case, what determines the "useful" bandwidth is also dependant on how well the complex gain curve can be calibrated by the digital hardware, and more importantly, how stable that calibration is.

The nominal mismatch in complex gain between two instances of the same filter design will obviously depend on the manufacturing tolerances involved, but in general will be proportional to the complex gain slope,

$$\Delta G \propto \frac{d}{df} s_{21}. \qquad (1)$$

Digital calibration in essence is a direct measurement of this deviation in complex gain between the I and Q channels as a function of frequency. The number of calibration points required is then proportional to the rate of change of this curve,

$$N \propto \left| \frac{d}{df} \Delta G \right| \propto \left| \frac{d^2}{df^2} s_{21} \right|. \qquad (2)$$

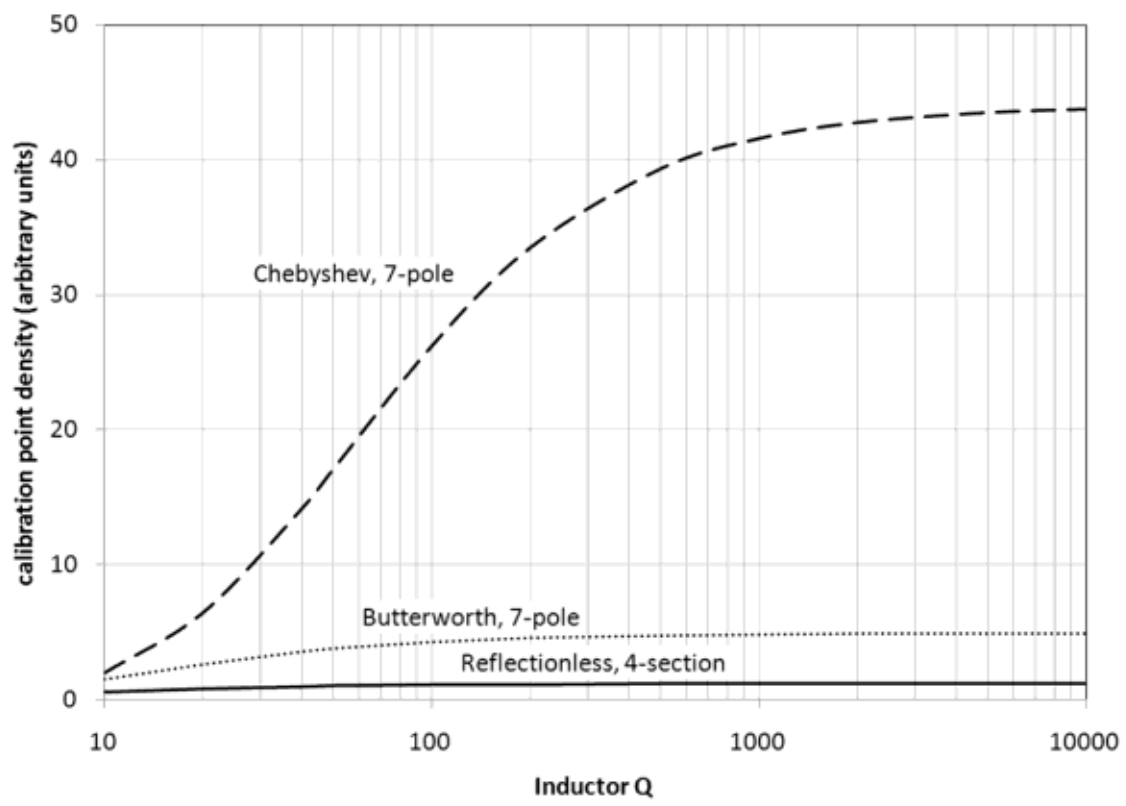

Figure 4. Relative calibration density required as a function of the inductor Q in the anti-aliasing filter.

The density of calibration points has an impact on the required signal processing overhead. The higher the calibration density, the more points at which the calibration will have to be sampled, the longer calibration will take, and the larger the memory that will have to be allocated to storing and recalling the table of coefficients. Further, depending on the spectrometer or correlator architecture, it may imply performing the FFT with a higher resolution than would otherwise be needed.

In general, the calibration stability is also proportional to the slope of the complex gain differential. Therefore, the calibration density may also be interpreted as a figure of merit for the calibration stability, where larger calibration density implies poorer stability.

The result of this calculation for each of the three filters is also plotted in Figure 2. The data has been normalized to the peak of the reflectionless filter calibration density curve. Although the insertion loss of the Chebyshev filter is smooth over the largest bandwidth, the phase is rapidly varying throughout its passband and particularly near the cutoff. Based on this result, the Chebyshev filter would typically require almost 15 times as many calibration points, or be 15 times less stable, as the reflectionless filter for the same level of sideband suppression. It is also reasonable to assume that the stability of the calibration with respect to temperature would be degraded by a similar factor. This information is summarized in Table 1.

Table 1
SUMMARY OF ANTI-ALIASING FILTER PERFORMANCE

| Filter Type | $BW_{3dB}$/ $BW_{Nyquist}$ | Insertion Loss (dB) | Cal. Density (arbitrary units) |
|---|---|---|---|
| Chebyshev (7-pole) | 0.85 | -1.4 | 14 |
| Butterworth (7-pole) | 0.71 | -1.2 | 4 |
| Reflectionless (4-section) | 0.74 | -0.6 | 1 |

Generally speaking, the required density of calibration points for a given filter type becomes larger as the filter order increases. This is shown for reflectionless filters with 1, 2, and 3 sections in Figure 3. Unlike conventional filters, the order of the reflectionless filter has little impact on the useful bandwidth or the slope of the cutoff, it only affects the peak of the out-of-band attenuation (approximately 15 dB per section).

Calibration density also becomes larger with increasing component Q. A plot of the peak calibration density as a function of inductor Q is shown in Figure 4. The reflectionless filter has a significant advantage over the other filter types in terms of the smoothness of its gain curve and, hence, upon the precision of the digital calibration as a function of frequency and temperature.

Arguably, the 3dB cutoff is not an appropriate definition of usable bandwidth in this application. Chebyshev and Butterworth filters are rarely if ever used beyond their 3 dB cutoff frequencies, not necessarily because of the excess loss, but because of the steepness of the gain slope at this point and the consequent sensitivity to manufacturing tolerances and temperature fluctuations. The reflectionless filter on the other hand has gentler complex gain slopes virtually everywhere -- gentler even in the cutoff region than the Chebyshev filter is in its passband -- and could be used as close to the Nyquist Limit as the system tolerance to aliased signals and noise will allow. The gain flatness within the usable band can be easily restored in digital signal processing. However, there is a subtle dynamic range trade-off in the sense that the gain-noise budget must be satisfied at the point of lowest analog gain in the system, but the large signal handling capacity must be satisfied at the highest gain point. This would be true for any band-limited receiver design.

Filter placement is another design tradeoff to consider. Putting the filters early in the gain sequence is preferable to protect the amplifiers from potentially strong out-of-band signals, while putting them in later keeps out-of-band noise from the post-filter amplifiers from aliasing into the spectrum. The unique reflectionless filters here have an advantage in that they are comprised of small, cascadable sections that need not be adjacent to one another. This makes it possible to distribute the filter sections throughout the IF path (see Figure 1), both before and after the IF amplifier, and providing good out-of-band padding to all the IF components, including the mixers, the amplifiers, and the ADCs.

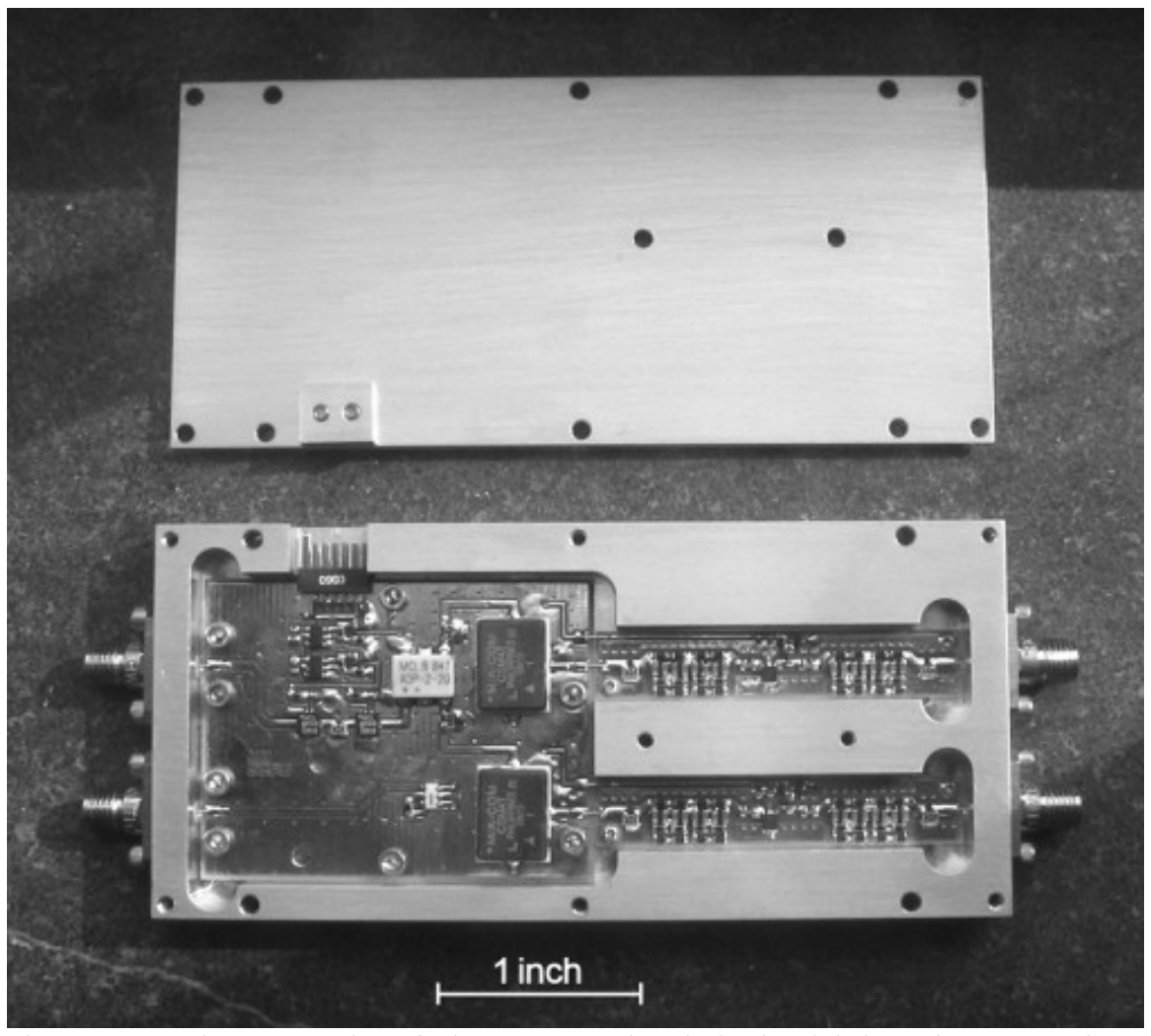

Figure 5. Photograph of the L-Band Digital Sideband-Separating Mixer prototype. The RF input (1200-1700 MHz) is on the upper left SMA connector, the LO input is on the lower left, and the I and Q output channels are on the right. Package dimensions are 4.25 x 2.0 x 0.5 in.

**4. Prototype Design**

A prototype DSSM was implemented in the form of an L-Band receiver capable of capturing a 500 MHz passband with a fixed LO at 1450 MHz. The analog portion is shown on the left side of Figure 1. The gain stages have been divided between RF and baseband frequencies. Not only is this preferred from a stability standpoint, it also effects a compromise between complex gain matching in the I and Q channels (which would favor more gain on the RF side) and dynamic range (which would favor more gain on the IF side). Note that we have split the IF filter into two sections as described above to provide some out-of-band attenuation between the mixers, amplifiers, and ADCs, and to reduce aliasing of out-of-band IF noise from the IF amplifier. For the purposes of this prototype, the Analog-to-Digital Converters were packaged separately, and the Digital Processing was performed in software after data acquisition.

A photograph of the completed DSSM module is shown in Figure 5. The module consists of three very small RF PCB's in a machined aluminum housing with SMA connectors for input and output. The components are all commercially-available surface mount ICs, and were selected particularly for gain flatness and impedance match. This helps to ensure that the complex gain parameters in each output channel vary slowly and smoothly with frequency, permitting a tight and stable calibration of the digital recombination coefficients at relatively few frequency points.

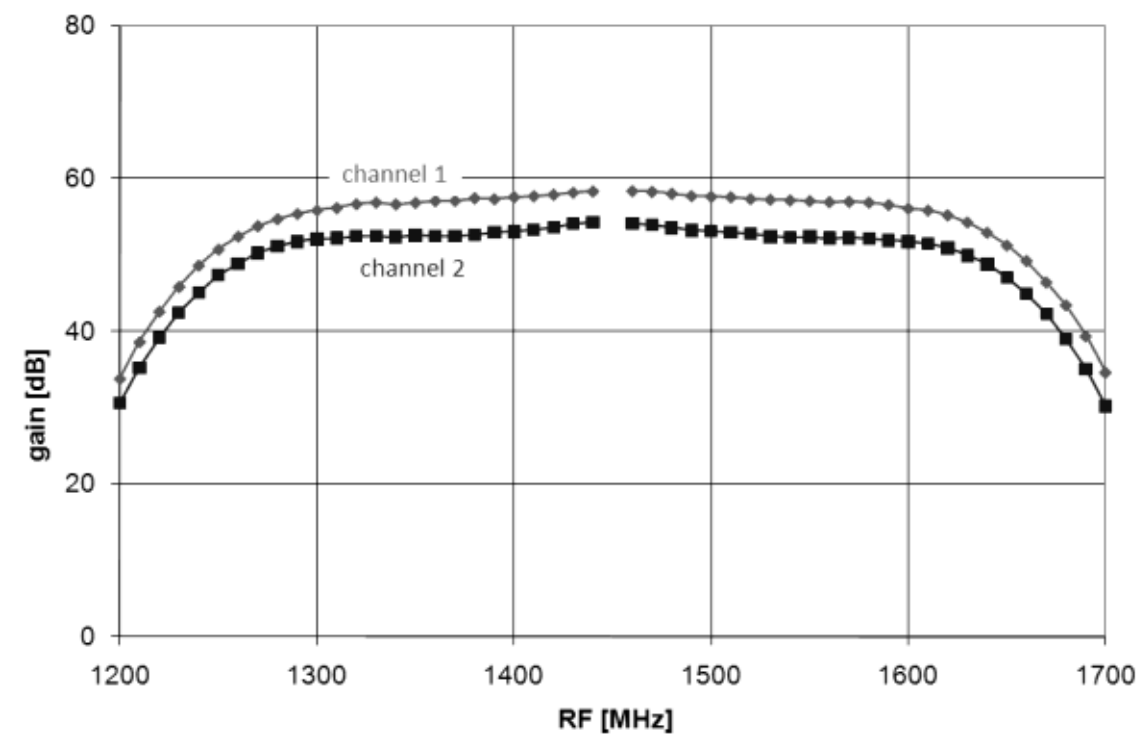

Figure 6. Measured intrinsic conversion gain of the L-Band downconverter (with no digital correction).

An earlier iteration of this design underscored the need for good isolation between the two output channels. Although a moderate level of crosstalk between the I and Q channels is, in principle, calibrated out by the digital processing, it introduces complex frequency structure into the amplitude and phase that force a larger number of calibration points to be used. Additionally, the more complex frequency structure can increase sensitivity to temperature fluctuations.

To improve the channel isolation, the IF components were placed on two narrow PCBs in separate cavities within the housing. This includes the IF gain stage and anti-aliasing filters. The IF boards were designed to be as narrow as possible, only 0.35 in. wide, so that a large number could be housed in parallel within minimal volume. For this prototype, however, only two boards were used, and their spacing was constrained by the output SMA launchers.

The raw (un-calibrated) passband of the analog downconverter is shown in Figure 6. It has a net conversion gain of about 55 dB (allowing another 30 dB for a cryogenic preamp during telescope tests). It draws about 160 mA from a +5V supply.

**5. Digital Calibration**

To provide optimum sideband isolation, the complex weighting coefficients must be calibrated to the particular amplitude and phase errors of the analog module. Calibration is performed by injecting a CW tone into one sideband and then sampling the detected outputs on the I and Q channels. Theoretically, any signal that is known to be in only one sideband can be used, even RFI, but it is convenient to provide a known calibration tone in the lab to ensure adequate signal-to-noise ratio. Our data will show that the weighting coefficients are stable enough to be done only once in a laboratory environment without having to be repeated in the field. However, if for some reason field calibration is desired, the calibration signal need not be injected in front of the first LNA. Any point before the RF signal splits to feed the two sub-mixers will do.

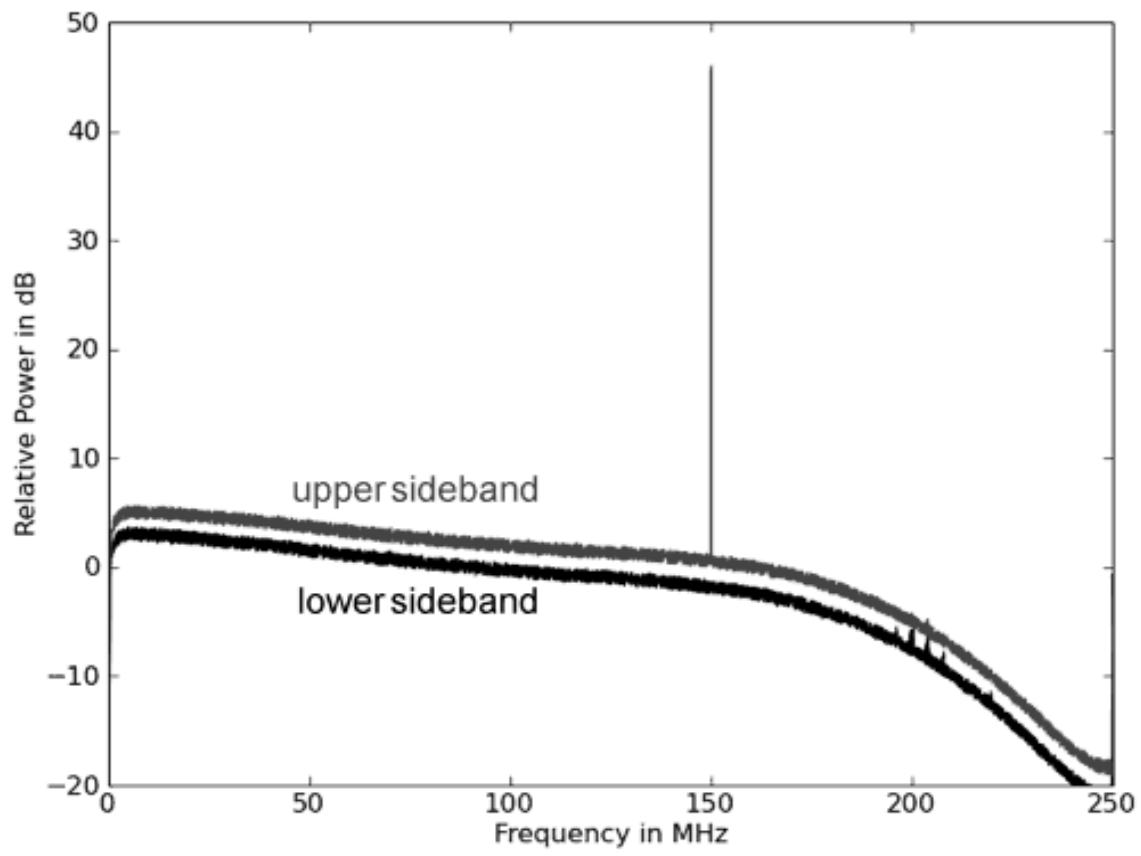

Figure 7. Plot of the measured power spectra after calibration and digital sideband separation. The curves are offset vertically for clarity. The strong tone at 150 MHz IF in the upper sideband is the calibration signal which has been completely suppressed in the lower sideband. The artifacts near 200 MHz are the result of digital noise in the sampler system.

As described in (Fisher & Morgan 2008), the gain amplitude ratio of the two channels is easily measured as the square root of the quotient of the time-averaged CW powers,

$$X = \sqrt{\left(V_1\overline{V}_1\right)/\left(V_2\overline{V}_2\right)}, \tag{3}$$

where $V_1$ and $V_2$ are the complex detected voltage amplitudes measured at the I and Q IF outputs, respectively. Similarly, the differential phase may be calculated as

$$\phi_{IF} \pm \phi_{LO} = \tan^{-1}\left(\mathrm{Im}\left(V_1\overline{V}_2\right)/\mathrm{Re}\left(V_1\overline{V}_2\right)\right) \tag{4}$$

where $\varphi_{IF}$ and $\varphi_{LO}$ are the IF-path and LO-path phase errors, respectively. The quantity, $\varphi_{IF}$ , includes any phase difference in the two signal paths caused by the RF power splitter before the mixers. The plus sign on the left of Eq. (4) applies when the calibration signal is in the lower sideband, and the minus sign applies when the calibration signal is in the upper sideband. Therefore,

$$\phi_{LSB} = \phi_{IF} + \phi_{LO} \tag{5}$$

$$\phi_{USB} = \phi_{IF} - \phi_{LO}. \tag{6}$$

It can be shown that the ratios of the weighting factors $C_1$ through $C_4$ in Figure 1 needed to cancel the lower and upper sidebands, respectively, are given by

$$\frac{C_1}{C_2} = \frac{1}{X} e^{-j(\phi_{LSB} - \pi)} \tag{7}$$

and

$$\frac{C_3}{C_4} = \frac{1}{X} e^{-j(\phi_{USB} - \pi)}. \tag{8}$$

The limiting sideband separation that can be expected due to uncertainty in the optimal calibration coefficients is determined by

$$S(dB) = 20\log_{10}\left(\phi_{err}\tfrac{\pi}{180}\right) - 6.02 \tag{9}$$

$$S(dB) = 20\log_{10}\left(X_{err} - 1\right) - 6.02. \tag{10}$$

Thus a 1° phase error will limit sideband suppression to -41 dB, while a 0.1 dB amplitude error will limit the suppression to -45 dB. Note that Equations (5) and (6) can be solved for $\varphi_{IF}$ and $\varphi_{LO}$, which may be useful in reducing the calibration task for a variable LO frequency by factoring the two phase terms.

## 6. Laboratory Tests

Initial tests of the L-Band prototype were performed in the lab. The fixed LO and RF calibration signals were generated by Agilent 8340B and 83640A synthesizers, respectively. The LO synthesizer was set to provide +13 dBm at 1450 MHz to the module, or about +10 dBm to each mixer. The RF was set to -60 dBm, with an external 30 dB pad, for a net RF input power level of -90 dBm, which was then stepped across the frequency band from 1200-1700 MHz. The analog outputs were connected to an external sampler through a pair of 5-foot long coaxial test cables. Although the cables were nominally the same length, no special care was taken to phase-match them. Finally, to test the stability of the calibration, the analog module was mounted on a hot plate with closed-loop temperature control.

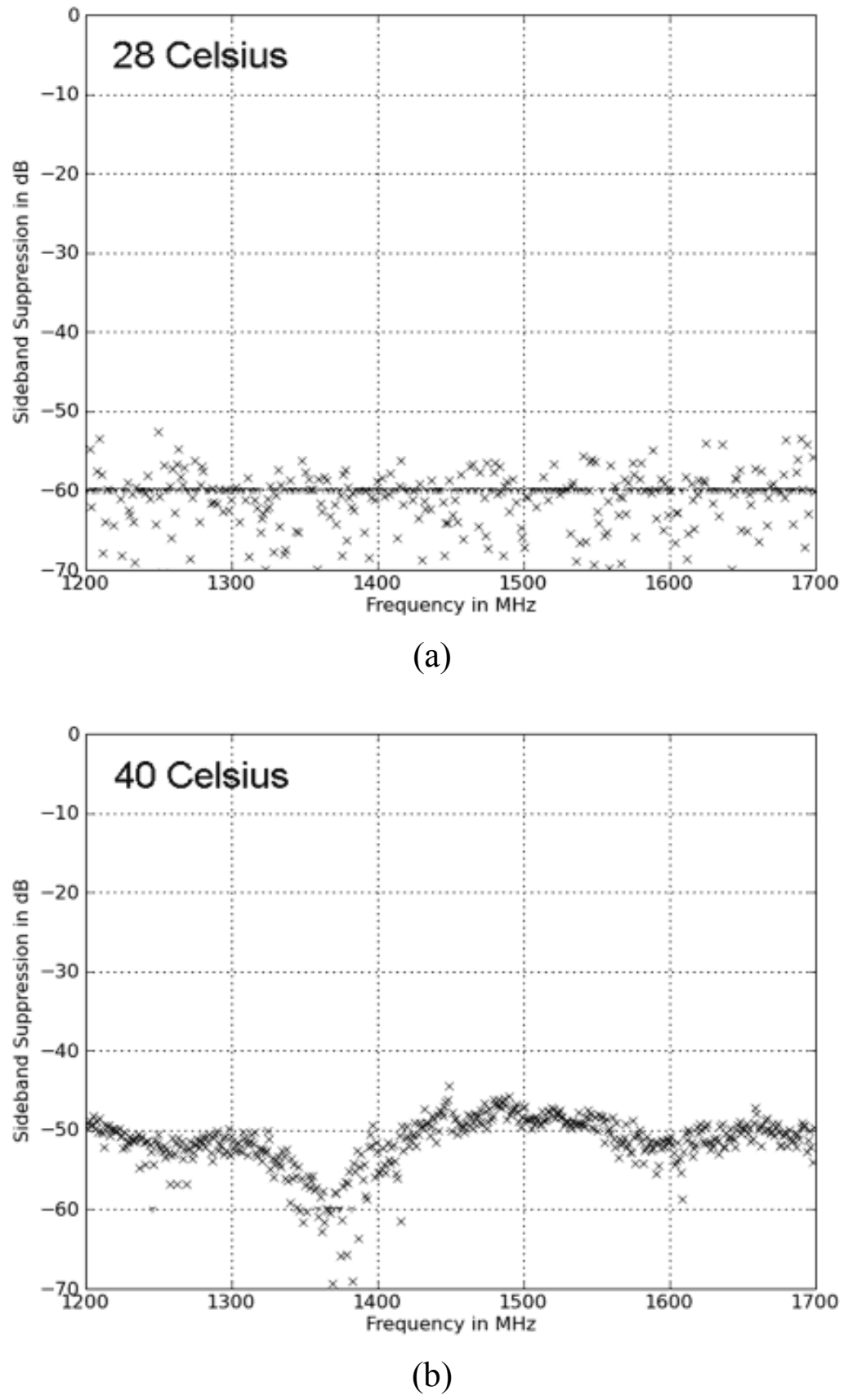

Figure 8. Sideband isolation measured in the lab at a) 28°C and b) 40°C. Points below the noise of the test set were conservatively marked at -60 dB as an upper bound. Calibration was performed at 28°C.

The analog-to-digital converter was a National Instruments Model PXI-5152, dual-channel, 8-bit, simultaneous sampling module running at its maximum rate of 500 MS/s. Data bursts were recorded in the sampler module memory for 130 ms and then archived to a PC hard disk. For the purposes of these experiments, calibration and sideband reconstruction were performed as a post-processing step in software. In actual astronomical use, this would be performed in real-time with a high-speed ASIC or FPGA. Overlapping, 65536-sample data windows were Fourier transformed, squared, and summed over the 130 ms burst to isolate the CW calibration signal with high signal to noise ratio.

A typical result is shown in Figure 7. In this 130 ms snapshot, a test tone at 150 MHz offset from the LO is preserved in the upper sideband spectrum, but virtually eliminated from the lower sideband spectrum. An investigation of the artifacts near 200

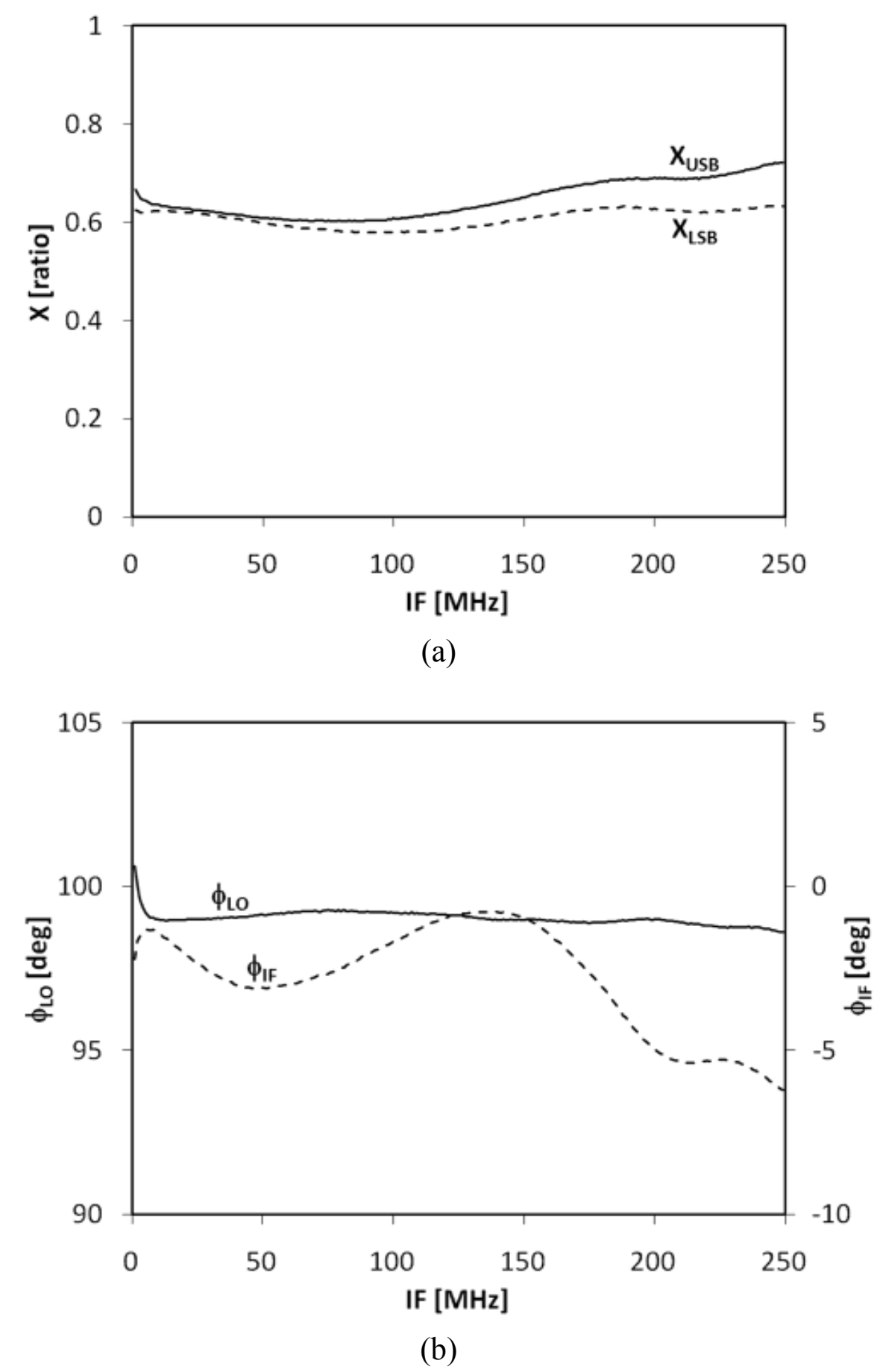

Figure 9. Measured calibration parameters of the mixer prototype.

MHz in the spectrum has shown that they result from digital noise in the sampler system and not from the analog mixer.

As summarized in Figure 8, sideband isolation is generally in excess of 60 dB across the full 500 MHz RF bandwidth, and degrades only to about 50 dB when the box temperature is raised by 12°C from its calibration temperature. This extreme stability of performance over a temperature range that far exceeds normal operational use indicates that calibration can be performed once in the lab after construction, avoiding the need to provide a calibration tone in the field.

The actual gain and phase terms measured during laboratory calibration are plotted in Figure 9. The nominal amplitude ratio, X, is about 0.6, or -4 dB. The local oscillator phase split, $\varphi_{LO}$, is about 99° instead of the ideal 90°. To first order, this parameter should be constant with IF, since the LO frequency is fixed, but small variations can occur due to imperfections on the RF side between the RF splitter and the mixers. The IF differential phase, $\varphi_{IF}$, varies between about 0° and -5°. All of these imperfections are removed as part of the calibration. What is important to note is that they are smooth and well behaved with frequency, indicating that relatively few calibration points may be needed with a simple interpolation algorithm. The exact density of calibration points required would depend on the permissible interpolation error and the resultant limit on image rejection as calculated using equations (9) and (10).

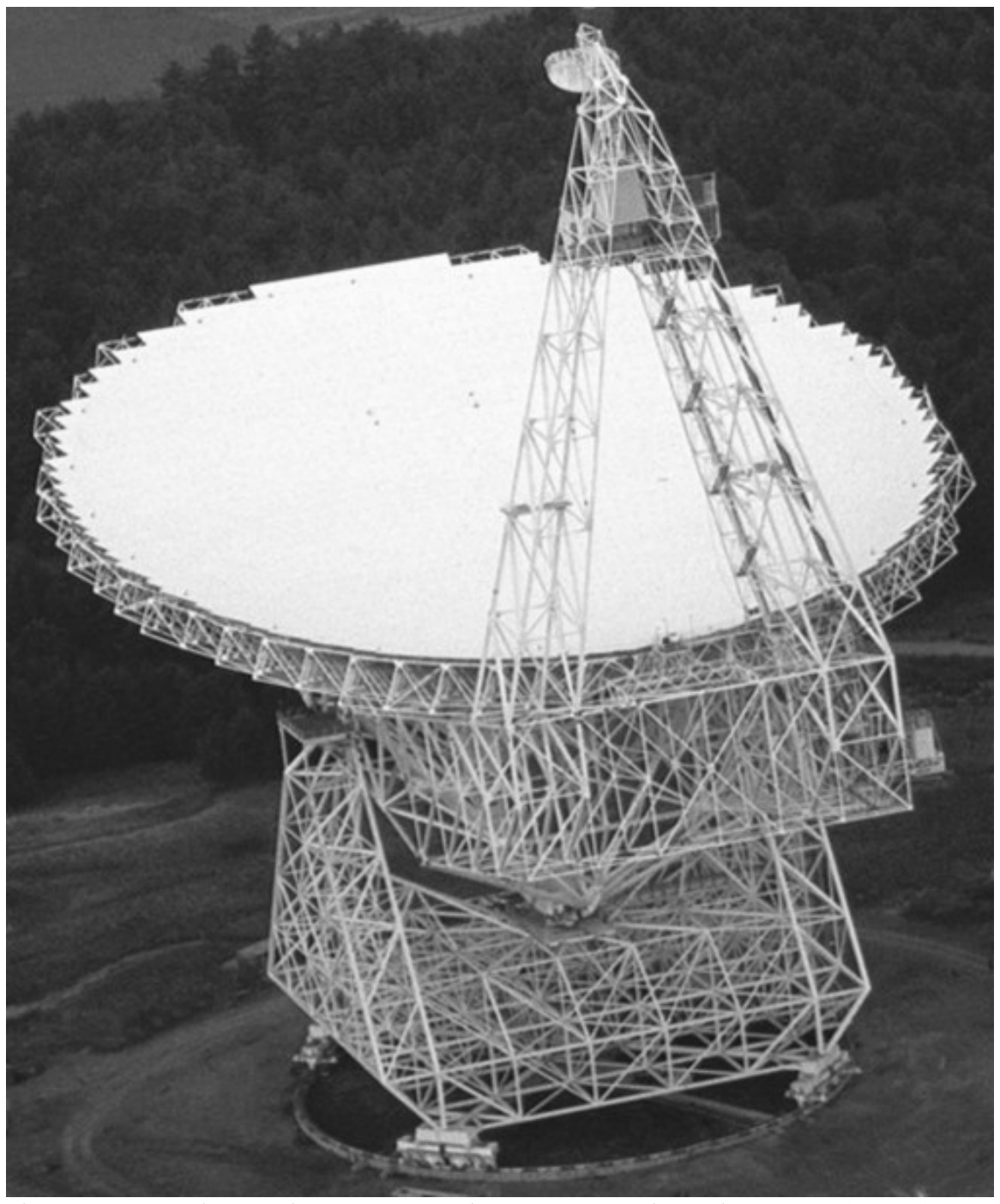

Figure 10. The Robert C. Byrd Green Bank Telescope (GBT) on which the prototype L-Band Digital Sideband Separating Mixer (DSSM) was tested.

The effect of digitizer quantization error on sideband separation was tested by software truncation of the low-order bits in the data samples. At least 60 dB of sideband separation was maintained as long as the effective input data word width was at least two bits (four levels). (The FFT complex coefficients are assumed to be accurate enough to maintain the required phase and amplitude accuracies stated above.) The most obvious effect of using very few bits for the input data samples is to scatter noise from the IF filter passband into its stop band due to intermodulation products caused by quantization errors (Thompson el al, 2001).

**7. Telescope Tests**

Following successful laboratory tests, the module was connected to the cold-stage output of the 1.2-1.8 GHz receiver on the Robert C. Byrd Green Bank Telescope (GBT, Figure 10) to verify its performance in the presence of Radio Frequency Interference (RFI). Although the telescope is located in the National Radio Quiet Zone (NRQZ) in West Virginia, this band is still heavily laden with RFI from a number of sources, including satellites (Geirland 2004).

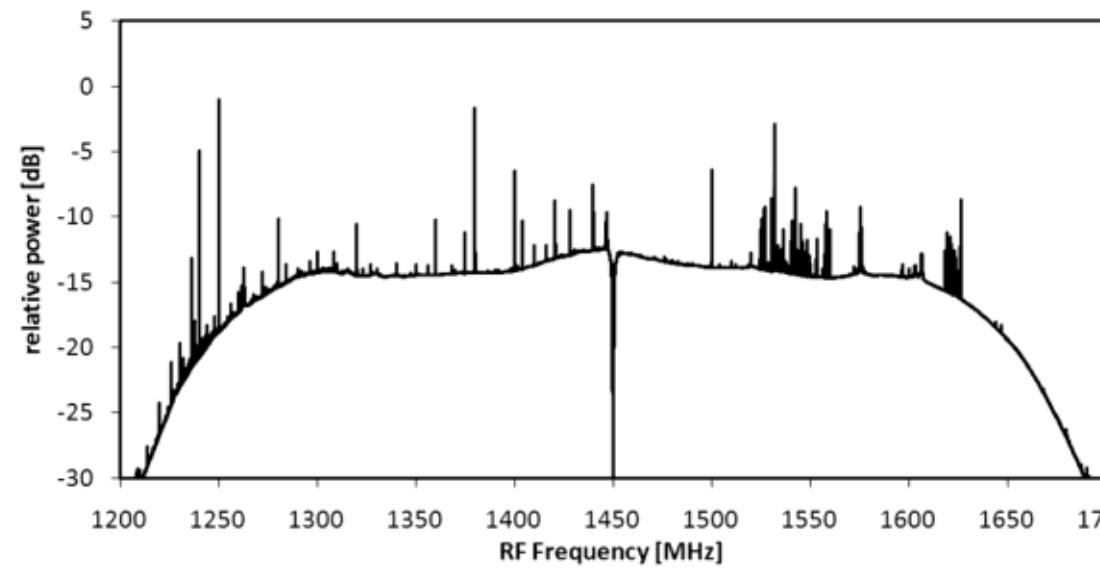

Figure 11. L-Band spectrum received using the DSSM on the GBT. Data shown is the average of about 500 snapshots of the complete spectrum for a total integration time of about 1 minute using a fixed LO at 1450 MHz.

Along with the downconverter module itself, test equipment was transported up to the receiver cabin of the telescope for the purposes of this experiment, including a synthesizer to generate the calibration tone, the dual-channel high-speed sampler, and the post-processing computer. As during our lab tests, data capture was limited to 130 ms snapshots. However, in order to increase sensitivity the data from 500 snapshots was averaged together to produce a single spectrum, shown in Figure 11, representing approximately 1 minute of integration time.

Because the test equipment was not permanently fixed to the structure, the telescope had to remain parked in the maintenance position, making it impractical to point the beam at any particular astronomical target. Since neutral hydrogen is present virtually everywhere in our galaxy, this celestial signal was present in all GBT data, as shown in Figure 12a. The remaining features are all believed to be external RFI, including for example the telemetry satellite band in the 1530-1547 MHz range, shown in Figure 12b. Careful comparison of the two sideband spectra shows no detectable remnant of any signal in the wrong sideband.

We can conclude from these tests that RFI is very effectively isolated to the sideband in which it originates, and that the presence of RFI in no way disrupts the accurate determination of calibration coefficients.

**8. The Zero-IF Hole**

One notable feature in the processed spectrum of Figure 11 that is worth discussing is the "hole" in the middle of the band where IF = 0 and the receiver has no sensitivity. This obviously arises from the AC-coupling of the IF path and may correctly be called a disadvantage of this design approach. It is trivial enough, however, to detune the LO frequency slightly, say by 5 MHz, should the astronomer wish to examine that portion of the spectrum. Such a test was in fact performed during the above measurements to ensure that all the peaks seen in the spectrum moved in the correct way, proving that they were true RF signals and not spurious tones leaking in at the IF frequency. In principle, digital calibration should be performed at both LO frequencies, but as was shown in Figure 8, the calibration terms are fairly smooth with frequency and the effect of a 5 MHz LO offset is quite small.

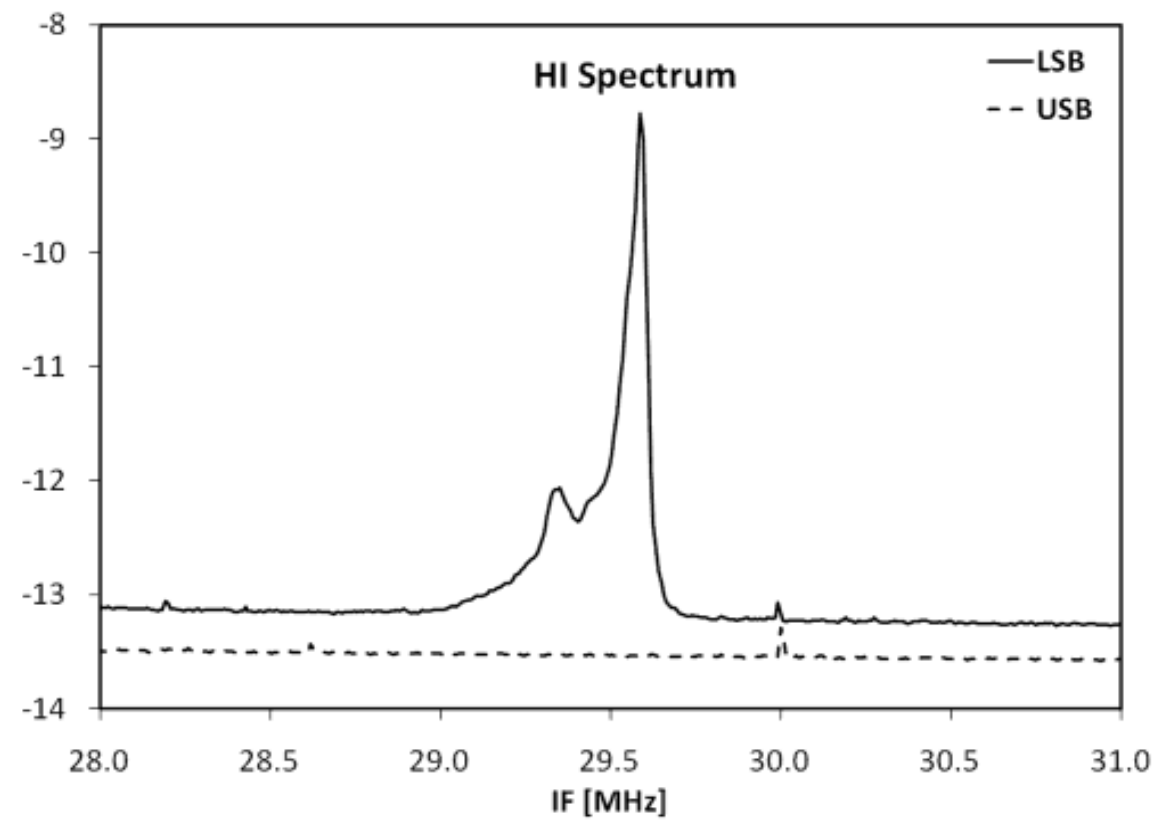

(a)

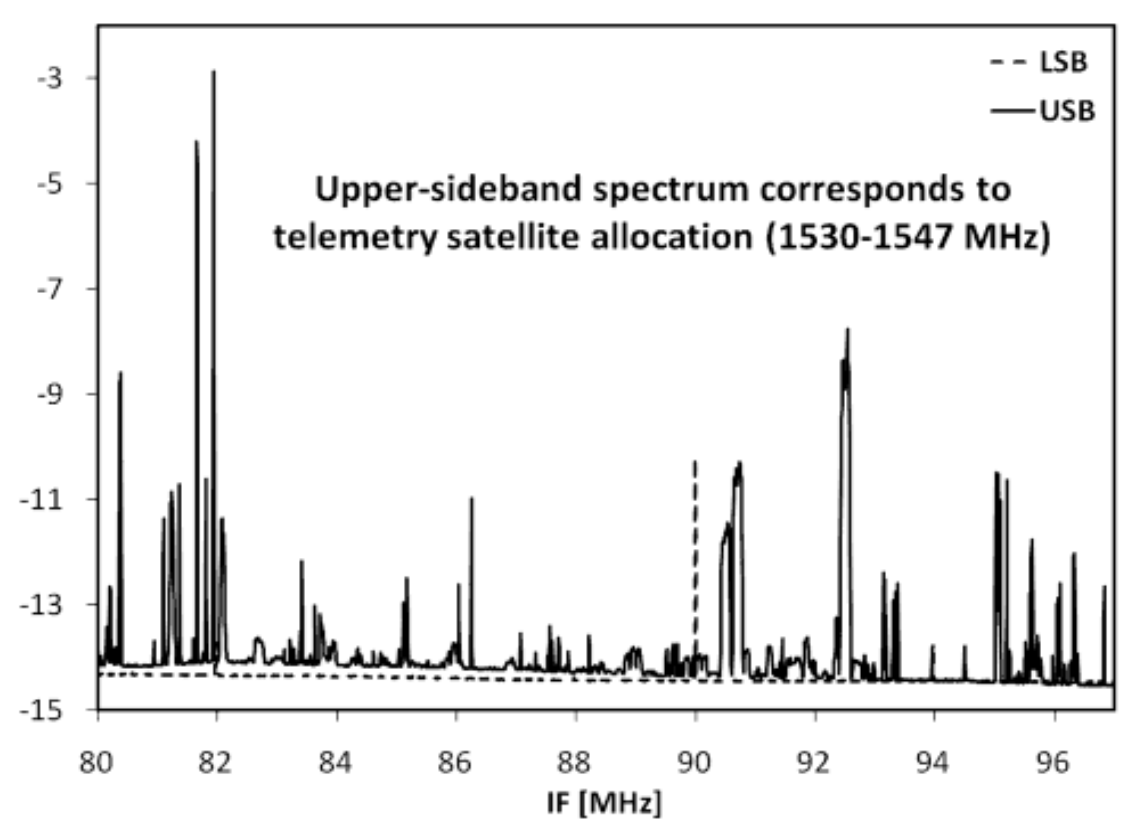

(b)

Figure 12. Detail view of portions of the L-Band spectrum. a) Galactic Hydrogen RF=1420.4 MHz, or IF=29.6 MHz in the lower sideband. b) Telemetry satellite band, 1530-1547 MHz, showing significant interference in the upper-sideband which is well-rejected in the lower-sideband. The vertical scale is relative dB.

The width of the hole in the prototype shown is about 1.2 MHz, meaning each channel operates down to an IF of 600 kHz. In this case, the limiting factor was the mixer, which is AC-coupled on its IF port. Another mixer could have been chosen, in which case the bias tee for the IF amplifiers would define the bandwidth at the low end. Large coupling capacitors with huge resonance-free bandwidths are readily available, but the RF choke inductors are more difficult. Still, with careful design it seems likely that IF chains with 10,000:1 bandwidth (say from 25 kHz to 250 MHz, for example) should be achievable in a straightforward manner.

DC coupling in the IF path is at least feasible, but would require a special amplifier in which the supply circuitry is isolated by topology rather than high-pass filtering. An operational amplifier, for example, would meet this criterion, but of course op-amps will be limited at high frequencies. Alternatively, one could have no IF amplifier at all, but that is unfavorable both in terms of dynamic range and stability as was discussed in the Design Tradeoffs section.

## 9. Conclusion

An integrated L-Band downconverter has been demonstrated which uses digital sideband reconstruction to achieve an extraordinary level of image-rejection with stable performance, better than 50 dB over a 12°C temperature change. This was the first step in a focused program to re-optimize radio astronomy receiver architecture to better take advantage of the latest advancements in digital technology. The downconverter described comprises the non-cryogenic part of the re-optimized receiver. Future work will include integrating the analog-to-digital converters in the package, as well as the conversion to an optical fiber signal. Other methods of simplifying the analog hardware in favor of digital hardware are being explored.

## Acknowledgments

The authors wish to thank Tod Boyd for assembly of the downconverter described, and their colleagues at the NRAO Central Development Laboratory for useful advice and discussions. This work was carried out at the National Radio Astronomy Observatory, a facility of the National Science Foundation operated under cooperative agreement by Associated Universities, Inc.